\documentclass[twocolumn,amsmath,amssymb,aps,prl]{revtex4-2}
\usepackage{CJK}
\usepackage{xcolor}
\usepackage{graphicx}
\usepackage{subfigure}
\usepackage{dcolumn}
\usepackage{bm}
\usepackage{tabularx}
\usepackage{amsmath}
\usepackage[colorlinks,linkcolor=blue, urlcolor=blue, anchorcolor=blue, citecolor=blue]{hyperref}
\usepackage[title]{appendix}

\usepackage{ulem}

\begin{document}
\title{Emergent Kagome lattice and non-Abelian lattice gauge field of biexcitons in t-MoTe\textsubscript{2}}
\author{Haochen Wang$^{1,2}$}
\author{Wang Yao$^{1,2}$}
\email{wangyao@hku.hk}
\address{$^1$ New Cornerstone Science Laboratory, Department of Physics, University of Hong Kong, Hong Kong, China}
\address{$^2$ HK Institute of Quantum Science and Technology, University of Hong Kong, Hong Kong, China}

\date{\today}
\begin{abstract}
Non-Abelian gauge fields, characterized by their non-commutative symmetry groups, shape physical laws from the Standard Model to emergent topological matter for quantum computation. Here we find that moir\'e exciton dimers (biexcitons) in twisted bilayer MoTe$_2$ are governed by a genuine non-Abelian lattice gauge field.
These dipolar-bound exciton dimers, formed on bonds of the honeycomb moir\'e superlattice, exhibit three quadrupole configurations organized into a Kagome lattice geometry, on which the valley-flip biexciton hoppings through electron-hole Coulomb exchange act as link variables of the non-Abelian lattice gauge theory.
The emergence of gauge structure here is a new possibility for composite particles, where the moiré electronic structure and interactions between the electron and hole constituents jointly enforce the underlying geometric constraint. The quadrupole nature of biexciton further makes possible local gate controls to isolate designated pathways from the extended lattice for exploiting consequences of non-commutative gauge structure including the genuine non-Abelian Aharonov-Bohm effect. This also provides a new approach for quantum manipulation of excitonic valley qubit. We show path interference on a simplest loop can deterministically transform the computational basis states into Bell states.
\end{abstract}
\maketitle

Gauge fields mediate nature's fundamental forces, governing interactions from subatomic to cosmological scales. In an Abelian gauge field, particles can acquire path-dependent phases that lead to the seminal Aharonov-Bohm (AB) effect. For non-Abelian fields, the behavior is markedly richer: even the order of paths matters due to non-commutativity~\cite{goldman_light-induced_2014}. Non-Abelian gauge field is central to a list of grand challenges in modern physics, from quantum chromodynamics to topological quantum computing. In artificial quantum systems, gauge fields can emerge through two primary mechanisms: (1) strong many-body correlations (e.g., fractional quantum Hall states) \cite{Lopez1991PRB}, or (2) geometric constraints on single particle dynamics (e.g. Bloch bands of lattice systems) \cite{NiuRMP2010,goldman_light-induced_2014}.
The latter has been exploited for synthesizing non-Abelian gauge fields in  laser-coupled cold atoms systems and optical systems~\cite{Osterloh2005PRL, Ruseckas2005PRL, wu_realization_2016, liang_chiral_2024,chen_non-abelian_2019, yang_synthesis_2019,cheng_non-abelian_2025}.
In particular, non-Abelian AB effect has been demonstrated using fiber optics~\cite{yang_synthesis_2019,cheng_non-abelian_2025}, in one-dimensional ladder of cold atoms~\cite{liang_chiral_2024}, and also simulated in electrical circuit~\cite{wu_non-abelian_2022}.

  
Moir\'e materials formed by two-dimensional transition metal dichalcogenides (TMDs) have provided a versatile platform for exploring condensed matter frontiers, from quantum matters of fundamental interest~\cite{shimazaki_strongly_2020,xu_correlated_2020,regan_mott_2020,huang_correlated_2021,park_dipole_2023,Eric2023Sci,cai_signatures_2023,zeng_thermodynamic_2023,park_observation_2023,XU2023prX, Richen2023Sci} to excitonic quantum optics and optoelectronic applications~\cite{yu_moire_2017,seyler_signatures_2019,tran_evidence_2019,jin_observation_2019,alexeev_resonantly_2019,brotons-gisbert_spinlayer_2020,li_dipolar_2020}.
TMDs moir\'e systems also make possible engineering emergent gauge fields for electrons, through exploiting the Berry phases arising from the moir\'e-patterned layer pseudospin texture~\cite{wu_topological_2019,yu_giant_2020,zhai_layer_2020,Dawei2020prm,zhai_ultrafast_2022}. 
In twisted MoTe$_2$ moir\'e superlattices, the Abelian gauge field in the adiabatic dynamics~\cite{yu_giant_2020} leads to the formation of flat Chern band in the ferromagnetic phase~\cite{Eric2023Sci}, underpinning the groundbreaking discovery of fractional quantum anomalous Hall effect at zero magnetic field~\cite{cai_signatures_2023,park_observation_2023,zeng_thermodynamic_2023,XU2023prX}. 
Non-Abelian gauge field can also emerge from non-adiabatic dynamics in inhomogeneously distorted moir\'e~\cite{zhai_layer_2020}, though probing its consequence exceeds existing capabilities. 
More fundamentally, for all such synthetic gauge field systems, a pivotal question persists:  
how can these engineered non-Abelian fields be harnessed for practical applications or fundamental advances?

In this work, we uncover a novel mechanism for the emergence of non-Abelian gauge fields in a composite particle, where the electronic structure of its constituents and their interactions jointly enforce the underlying geometric constraint. The particle concerned is a bound dimer of moir\'e excitons (i.e., a biexciton) in twisted bilayer MoTe$_2$, formed by excitonic dipole attraction on the nearest-neighbor bonds of the honeycomb moir\'e superlattice. Such biexcitons are thermodynamically favored over a broad range of intermediate temperatures, and have three quadrupole configurations, in addition to the valley pseudospins. 
We show that electron-hole Coulomb exchange enables valley-flip biexciton hoppings, which organize into a Kagome lattice geometry. These hoppings act as $U(2) \times U(2)$ link variables in a genuine non-Abelian lattice gauge theory, where the two $U(2)$ groups operate respectively on the valley pseudospins of the two individual excitons within a dimer.
We find exotic nodal-ring zero-modes in the biexciton band dispersion, where the effective mass diverges along high-symmetry lines and vanishes orthogonally.
The quadrupole nature of biexciton further makes possible local gate control to isolate simple pathways from the extended lattice for exploiting the intriguing consequences of the non-commutative gauge structure including the genuine non-Abelian AB effect.
Remarkably, we find that
non-Abelian AB interference on simplest loops can realize entanglement gate for the valley pseudospin qubits, and deterministically
transform the computational basis states
into Bell states.

\begin{figure*} [htb]\centering
\includegraphics[width=0.65\textwidth]{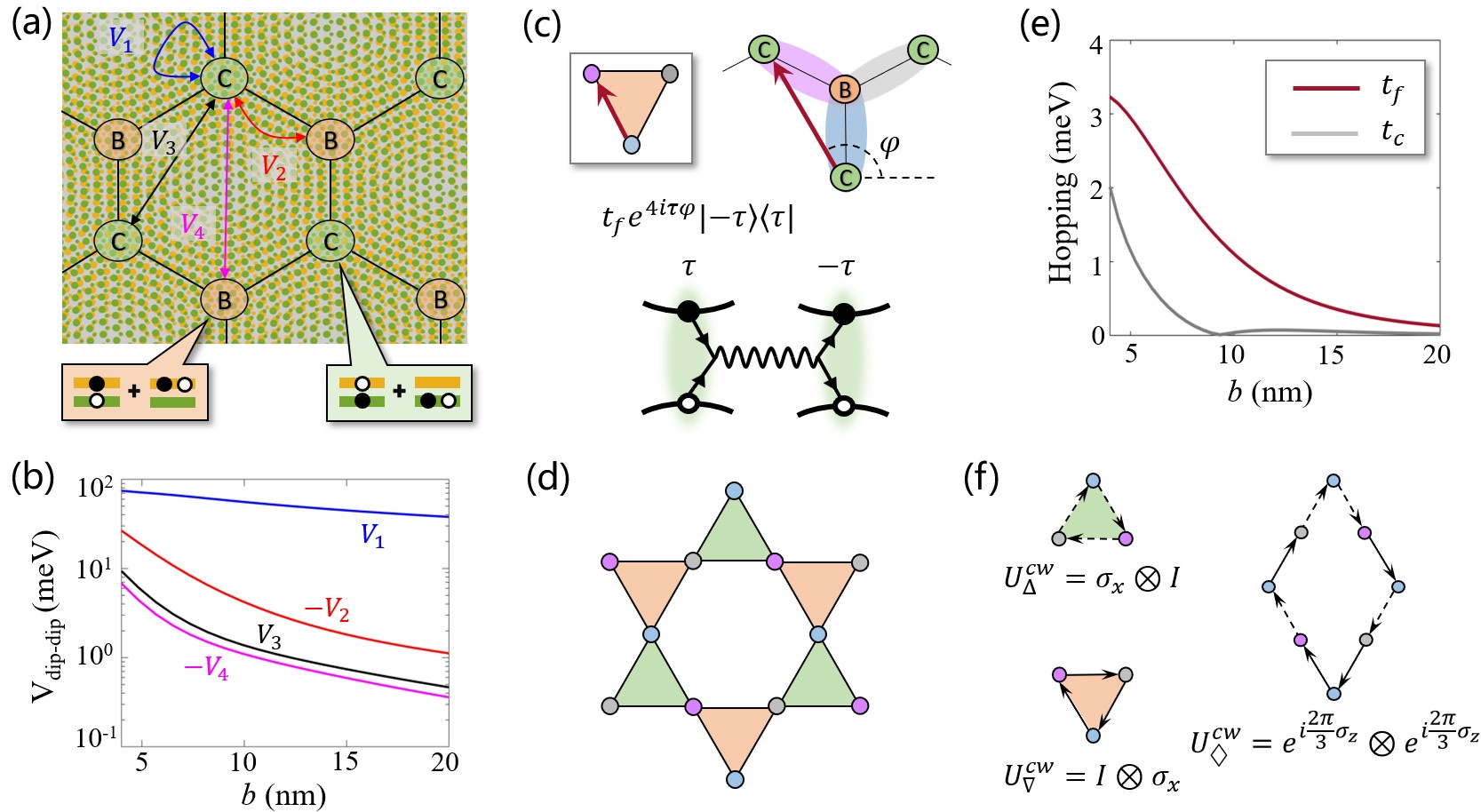}
\caption{(a) Schematic of twisted MoTe\textsubscript{2} moir\'e superlattice where hybrid excitons are trapped at B and C sites with opposite electric dipoles. (b) Dipole-dipole interactions between hybrid moir\'e excitons. (c) Biexcitons formed on nearest-neighbor bonds by dipole attraction ($V_{2}$). Red arrow denotes the effective hopping between biexciton configurations, due to the valley-flip electron-hole exchange. Inset shows the equivalent biexciton representation, where color coded dots denote biexcitons of distinct electric quadrupole. (d) Emergent Kagome lattice in the biexciton representation. Three biexciton states on each brown (green) triangle share a common B (C) site. (e) Valley-flip ($t_{f}$) vs. valley-conserved ($t_{c}$) biexciton hopping as functions of moir\'e period $b$. The weight of the intralayer component is taken to be $\frac{1}{6}$ (c.f. Supplementary Note 1). (f) Three loop operators under the dominant valley-flip hopping. Evidently $[U_{\bigtriangledown},U_{\diamondsuit}]\neq0$, i.e. the lattice gauge field is genuinely non-Abelian.}
 \label{Fig1}
\end{figure*}


Twisted MoTe$_2$ bilayer of near 0 degree twisting features interlayer electrical polarization locked to the local stacking registry, which is spatially varying 
with the moir\'e periodicity~\cite{weston_interfacial_2022,wang_interfacial_2022}. Dipolar interlayer excitons are trapped - by such background electrical polarization - at the MX and XM stacking regions with opposite
layer configuration (electric dipole), denoted as B and C sites respectively in Fig.~\ref{Fig1}a. The trapping energy can largely compensate the binding energy difference from an intralayer exciton. This leads to the formation of hybrid moir\'e excitons at the honeycomb superlattice sites, where the hybridization ratio can be tailored through environmental dielectric engineering that tunes the binding energy difference between inter- and intra-layer components. The $C_3$ rotational symmetry at the trapping sites, on the other hand, dictates that the hybridized intralayer component has a $p$-type center-of-mass wavefunction, where destructive interference leads to a vanishingly small optical transition dipole~\cite{zheng2024forster}. 

\begin{figure} [htpb]\centering
\includegraphics[width=0.5\textwidth]{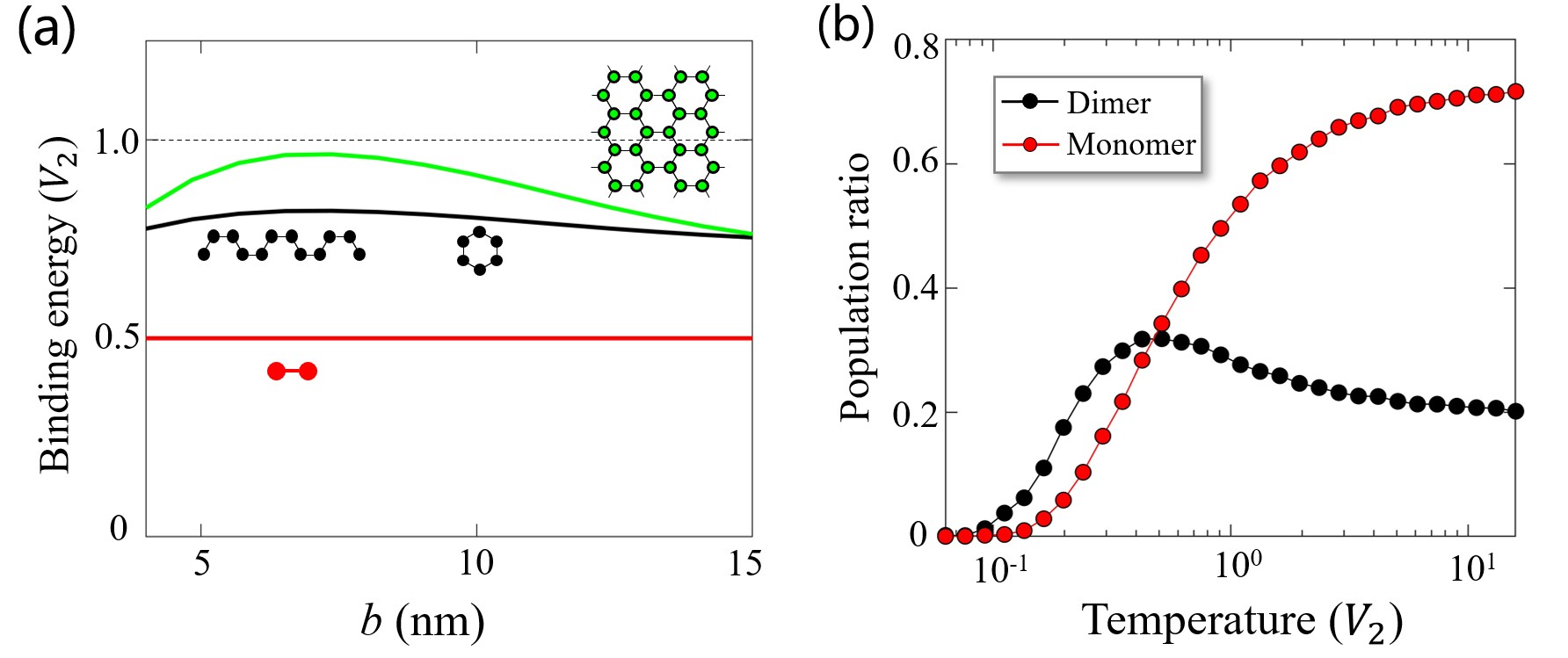}
\caption{(a) Binding energy per exciton vs moir\'e period b, in different dipolar bound configurations. Green denotes extended cluster (area$\gg$boundary), black for armchair line or hexagon, and red for dimer (biexciton). (b) Population ratio of moir\'e excitons that exist in dimer and monomer forms respectively, as function of temperature at a filling of $0.1$ exciton per cell. $b = 11$ nm where the energy/temperature unit $V_{2}\sim 3$ meV or $35$ K (c.f. Fig.~\ref{Fig1}b). See Supplementary Note 3 for details of Monte Carlo simulation.}
 \label{Fig2}
\end{figure}

Such hybrid moir\'e excitons are thus symmetry protected to have long radiative lifetime, even with a significant intralayer component~\cite{Hongyi2021PRX,zheng2024forster}. Through the electric dipole of the interlayer component, they interact repulsively within each sublattice and attractively between the two sublattices. 
Fig.~\ref{Fig1}b plots the strength of four leading interaction channels as function of moir\'e period $b$ (c.f. Supplementary Note 2). On-site repulsion $V_{1}$ in the order of tens of meV effectively excludes double occupation at low exciton temperature. The dominant interaction channel becomes the nearest-neighbor (NN) attraction $V_{2}$, which binds excitons into dimers, or a biexciton that carries electric quadrupole moment on the NN bond. Such biexcitons have a sizable binding energy $\sim10$ meV at a moir\'e period of $\sim 10$ nm (Fig.~\ref{Fig1}b). Excitons can further segregate into larger clusters, which are also affected by the much weaker next-NN repulsion $V_3$ and next-next-NN attraction $V_4$, and so on (c.f. Supplementary Note 3). The additional energy gain by forming larger clusters is not as significant as that upon forming dimers, as shown by plots of the binding energy per exciton in Fig.~\ref{Fig2}a. This means that, at finite temperature where the exciton clusters dissociate by thermal fluctuations, biexcitons can be favored. Fig.~\ref{Fig2}b shows the ratio of moir\'e hybrid excitons that exist in dimer form at a filling of 0.1 exciton per cell, from Monte Carlo simulations. Biexciton ratio exceeds $30 \%$ at 20 K, and remains significant up to room temperature. We focus on such quadrupolar biexcitons, which are of clear experimental relevance in a thermal gas over broad range of intermediate temperatures, and in the dilute limit down to low temperature.

\textit{Biexciton hopping on the effective Kagome lattice} -- These biexciton exhibits three distinct quadrupole configurations determined by the orientation of NN bonds, as highlighted by the oval shades in Fig.~\ref{Fig1}c. Two biexciton configurations sharing a common B or C site can be inter-converted through a single step hopping of one exciton in the dimer. This enables the biexciton to hop in the moir\'e superlattice as easily as an individual exciton, whereas the spatial configurations and their connections constitute an effective Kagome lattice, with three orbitals of distinct quadrupoles denoted as $D_{1}$, $D_{2}$, and $D_{3}$. Each orbital also has an internal degree of freedom spanned by $\{|\tau\rangle_{B}|\tau'\rangle_{C}$\}, where 
$\tau,\tau'=\pm$ are the excitonic valley pseudospins at B and C site, optically accessible through the valley selection rule~\cite{Hongyi2021PRX}.
Note this biexciton Kagome lattice only has nearest-neighbor hopping: any direct further-neighbor hopping is a second-order process through a far-detuned two-exciton configuration, therefore negligibly small.    

Generally, hopping of hybrid excitons in the moir\'e takes place through two microscopic mechanisms~\cite{zheng2024forster}: (i) the valley-conserving kinetic propagation in the moir\'e potential~\cite{Hongyi2017scia}; (ii) the electron-hole Coulomb exchange, which can annihilate a $\tau$ valley exciton at one moir\'e site and non-locally create one in valley $\tau'$ at another site (c.f. Fig.~\ref{Fig1}c). The latter is essentially the F\"orster coupling well known as an energy transfer mechanism~\cite{yu_dirac_2014,Li_2024,ha_dual_2014,forster_energiewanderung_1946,hichri_resonance_2021}. Here it plays the role of a coherent hopping channel between the sites of moir\'e superlattice, which is activated only when the exciton has an intralayer component. Such F\"orster coupling channel for the hybrid moir\'e excitons in twisted MoTe$_2$ has been systematically analyzed in
Ref.~\cite{zheng2024forster}. For the biexcitons here, the relevant channels are the valley-flip and valley-conserved hopping between a closest pair of sites both on B or on C sublattice (c.f. Fig.~\ref{Fig1}c). 
In Fig.~\ref{Fig1}e, we compare the valley-flip and valley-conserved hopping strengths as functions of moir\'e period. Notably, the valley-flip F\"orster coupling has a longer range as compared to the valley-conserved one, the latter further subject to destructive interference with the kinentic propagation (see Supplementary Note 1). As a result, the valley-flip process dominates the biexciton hopping on the Kagome lattice, which takes the form: $t_f (e^{4i \varphi} | - \rangle \langle + | + e^{-4i \tau \varphi} | + \rangle \langle - |)$, $\varphi$ denoting the angle of the hopping direction.




\textit{Non-Abelian lattice gauge field} -- Biexcitons in t-MoTe\textsubscript{2} is therefore described by the low-energy effective Hamiltonian:
\begin{eqnarray}
\hat{H}&=&\sum_{\vec{r}}t_{f}(\hat{b}^{\dagger}_{1,\vec{r}}\hat{h}_{1}\hat{b}_{2,\vec{r}}+\hat{b}^{\dagger}_{2,\vec{r}}\hat{h}_{2}\hat{b}_{3,\vec{r}}+\hat{b}^{\dagger}_{3,\vec{r}}\hat{h}_{3}\hat{b}_{1,\vec{r}})\nonumber\\
&+&\sum_{\vec{r}}t_{f}(\hat{b}^{\dagger}_{1,\vec{r}}\hat{h}'_{1}\hat{b}_{2,\vec{r}+\vec{\delta_{1}}}+\hat{b}^{\dagger}_{2,\vec{r}}\hat{h}'_{2}\hat{b}_{3,\vec{r}+\vec{\delta_{2}}}\nonumber\\
&+&\hat{b}^{\dagger}_{3,\vec{r}}\hat{h}'_{3}\hat{b}_{1,\vec{r}+\vec{\delta_{3}}})+h.c.
\label{eq1}
\end{eqnarray}
Here $\hat{b}^{\dagger}_{1,\vec{r}}$, $\hat{b}^{\dagger}_{2,\vec{r}}$, and $\hat{b}^{\dagger}_{3,\vec{r}}$ are the creation operator of the $D_{1}$, $D_{2}$, and $D_{3}$ biexciton at the Kagome unit cell centered at $\vec{r}$.  $\vec{\delta}_{i}$ are primitive lattice vectors, related by $C_{3}$ rotation. $\hat{h}_{j} \equiv \hat{I}\otimes(\hat{\sigma}_{x} \cos\varphi_{j} + \hat{\sigma}_{y} \sin\varphi_{j})$ and $\hat{h}'_{j} \equiv ( \hat{\sigma}_{x} \cos\varphi_{j} + \hat{\sigma}_{y} \sin\varphi_{j})\otimes \hat{I}$ are the valley-flip operators upon the biexciton hopping, acting respectively on dipole up exciton (C site) and dipole down exciton (B site). $\varphi_{j}=0,\frac{2\pi}{3},-\frac{2\pi}{3}$ are the angles of hopping directions $D_{1}\to D_{2}$, $D_{2}\to D_{3}$, and $D_{3}\to D_{1}$, respectively. Valley-conserved hopping is neglected here because of its small magnitude, especially for 
moir\'e period $b \geq 10$ nm
(c.f. Fig.~\ref{Fig1}e). 

When $\varphi_j\neq \varphi_{j'}$, $[\hat{h}_{j},\hat{h}_{j'}]\neq 0$ and $[\hat{h}'_{j}, \hat{h}'_{j'}]\neq 0$, namely the valley rotations upon the hoppings are non-commutative. 
The set of $\hat{h}_{j}$ and $\hat{h}'_j$ are the elements of a $U(2)\times U(2)$ group, and the hopping processes in Eq.~(\ref{eq1}) act as link variables for a corresponding non-Abelian lattice gauge theory. 
The non-Abelian properties are illustrated by three simplest loop paths in Fig.~\ref{Fig1}f. The loop operators are given by $U^{CW}_{\bigtriangleup}=\hat{\sigma}_{x}\otimes I$, $U^{CW}_{\bigtriangledown}=I\otimes \hat{\sigma}_{x}$, and $U^{CW}_{\diamondsuit}=e^{i\frac{2\pi}{3} \hat{\sigma}_{z}}\otimes e^{i\frac{2\pi}{3} \hat{\sigma}_{z}}$, which realize valley flips of dipole down and dipole up exciton, and valley-dependent phase shift, respectively.
The non-commutativity $[U_{\bigtriangledown},U_{\diamondsuit}]\neq0$ characterize the genuine non-Abelian nature of this $U(2)\times U(2)$ lattice gauge field \cite{goldman_light-induced_2014}.

\begin{figure} [htb]\centering
\includegraphics[width=0.38\textwidth]{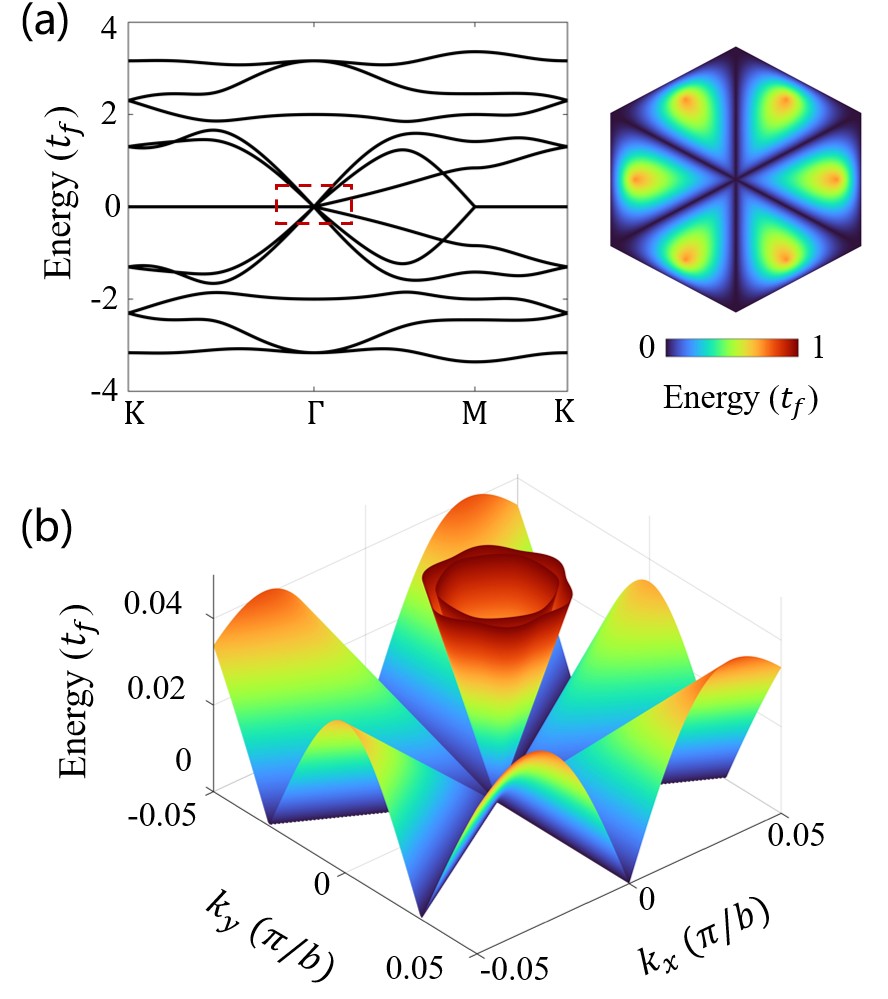}
\caption{(a) Biexciton bands in the Kagome lattice, featuring doubly degenerate nodal-ring zero modes along the high symmetry lines $\bar{K}-\Gamma-K$ and $\bar{K}-M-K$. Energy is in unit of $t_{f}$, the valley-flip hopping amplitude. (b) Dispersion near the $\Gamma$ point (c.f. dashed box in (a)).}
 \label{Fig3}
\end{figure}

\textit{Biexciton dispersion} -- 
Fig.~\ref{Fig3} presents the dispersion of biexcitons under the non-Abelian lattice gauge field in Eq.~(\ref{eq1}).
There exist doubly degenerate nodal-ring zero modes, where biexciton effective mass diverges along the high-symmetry lines $\bar{K}-\Gamma-K$ and $\bar{K}-M-K$, and vanishes orthogonally with a Dirac-like dispersion. The zero-lines cross at $\Gamma$ point with a six-fold degeneracy, where two additional Dirac cones emerge (Fig.~\ref{Fig3}b). The gauge field here preserves three symmetries: (i) time-reversal symmetry; (ii) particle-hole symmetry; and (iii) chiral symmetry. The $U(2)\times U(2)$ lattice gauge structure underlies the form of Hamiltonian having off-diagonal blocks only, and the rank of the off-diagonal block determines the existence and degeneracy of nodal-ring zero modes along high-symmetry lines~\cite{yuxin2024arxiv}. 
See details in Supplementary Note 4. 

\begin{figure} [htb]\centering
\includegraphics[width=0.4\textwidth]{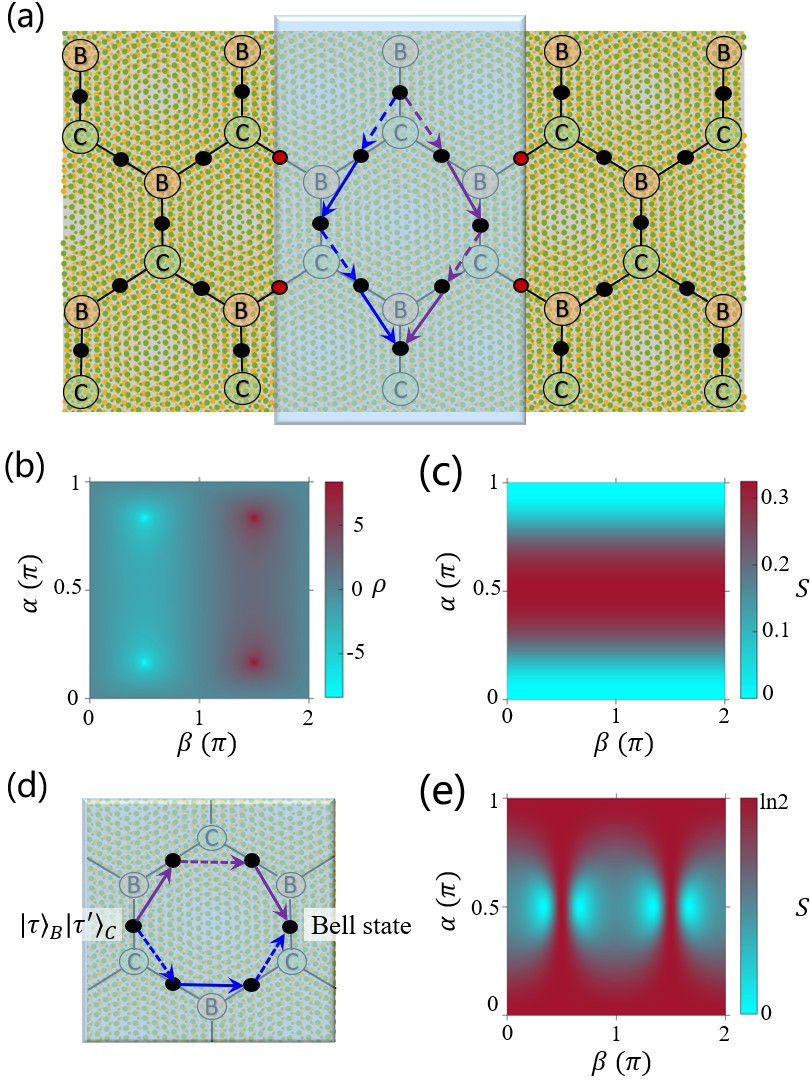}
\caption{(a) Local gate control for effective isolation of pathways for the non-abelian Aharonov-Bohm effect. Because of the quadrupole moment of biexciton, Stark shift by the interlayer bias occurs only at boundary of the gated area (red dots). By such detuning, biexciton lattice sites inside can be decoupled from outside. (b,c) Non-abelian AB interference of evolutions along the two symmetric pathways in (a) with initial state $|\vec{n}\rangle_{B}|\vec{n}\rangle_{C}$. $\rho=\frac{\langle\leftarrow\leftarrow|\Psi_{f}\rangle}{\langle\rightarrow \rightarrow|\Psi_{f}\rangle}$ and entanglement entropy $S$ are plotted in (b) and (c) respectively for the path-interfered final state, as functions of Bloch sphere angles of the initial state $|\vec{n}\rangle=[\cos\frac{\alpha}{2},\sin\frac{\alpha}{2}e^{i\beta}]^{T}$. (d) Non-abelian AB interference around a hexagon turns an initial state $|\tau\rangle_{B}|\tau'\rangle_{C}$ to the pure Bell state $e^{i\tau'\frac{2\pi}{3}}|-\tau\rangle_{B}|\tau'\rangle_{C}+e^{-i\tau\frac{2\pi}{3}}|\tau\rangle_{B}|-\tau'\rangle_{C}$. Two Bell states can therefore be deterministically realized. (e) Entanglement entropy of the final state upon the path interference in (d), for the same sets of initial states as in (b,c).}
 \label{Fig4}
\end{figure}

\textit{Non-Abelian AB effect and Bell state production under local gate control} -- Electric gate control of biexciton is made possible through its quadrupole moment, which can couple to the lateral gradient of a perpendicular electric field. Such field gradient can be naturally realized at the boundary of a top or bottom gate, as schematically illustrated in Fig.~\ref{Fig4}a. On honeycomb bonds either inside or outside the gated area, the biexciton energy does not respond to the interlayer bias, as Stark shift of its B and C components cancel each other. On partially covered bonds at the boundary (c.f. red dots in Fig.~\ref{Fig4}a), the biexciton energy can acquire a Stark shift in the order of the interlayer bias. Either a positive or negative energy shift $ \gg t_f$ effectively decouples the affected lattice sites. A modest local gate bias can therefore be implemented for isolating designated area from the rest of the lattice. 

The gated area illustrated in Fig.~\ref{Fig4}a encloses a diamond-shaped interference pathway, for which we examine the AB effect as well as valley pseudospin manipulations in the non-Abelian lattice gauge field. For biexciton in a valley basis state $|\tau\rangle_{B}|\tau'\rangle_{C}$, the evolution along the blue and purple pathways with even number of valley flips  will end up in the same state $|\tau\rangle_{B}|\tau'\rangle_{C}$ with a phase change of $e^{i(\tau+ \tau')\frac{2\pi}{3}}$ and $e^{-i(\tau+ \tau')\frac{2\pi}{3}}$ respectively. The 
$\tau, \tau'$ dependence of the phase leads to path-dependent valley rotations about the $z$-axis for a general initial state, which is the characteristics of non-Abelian AB effect~\cite{goldman_light-induced_2014,liang_chiral_2024,dong_temporal_2024}. Consider an initial state $|\vec{n}\rangle_{B}|\vec{n}\rangle_{C}$, where $\vec{n}$ is a general valley Bloch vector common for both B and C components. 
Along either the blue or purple path, the propagation in the gauge field preserves this form of wavefunction, leading to a common valley rotation for B and C components with path-dependence. This results in a final state of the form $|\Psi_{f}\rangle = e^{-i\frac{2\pi}{3}}|\vec{n}_1\rangle_{B}|\vec{n}_1\rangle_{C} + e^{i\frac{2\pi}{3}}|\vec{n}_2\rangle_{B}|\vec{n}_2\rangle_{C}$.  
The final state is characterized in Fig.~\ref{Fig4}b and \ref{Fig4}c in terms of valley polarization ratio $\rho=\frac{\langle\leftarrow \leftarrow|\Psi_{f}\rangle}{\langle\rightarrow \rightarrow|\Psi_{f}\rangle}$ and entanglement entropy $S$, as function of the sphere angles of initial Bloch vector $\vec{n}$. Characteristics of non-Abelian AB effect are clearly seen here~\cite{liang_chiral_2024,dong_temporal_2024}. 

The non-Abelian AB interference in the local-gate defined pathway can therefore be implemented for entanglement generation between the valley pseudospins of the exciton components, which can serve as qubit carriers given their long radiative lifetime. In Fig.~\ref{Fig4}d, we 
exploit interference pathway on a single hexagon unit, for generation of maximum entanglement. Consider an input on any of the basis states $|\tau\rangle_{B}|\tau'\rangle_{C}$, upon the propagation along the blue and purple paths, the final state of interference become: $e^{i\tau'\frac{2\pi}{3}}|-\tau\rangle_{B}|\tau'\rangle_{C}+e^{-i\tau\frac{2\pi}{3}}|\tau\rangle_{B}|-\tau'\rangle_{C}$, which is always a Bell state. 
Fig.~\ref{Fig4}e plots the entanglement entropy of the final state upon the path interference for a general initial state $|\vec{n}\rangle_{B}|\vec{n}\rangle_{C}$, where maximal entanglement ($S=\ln2$) is achieved over a broad region of the initial state parameter space. This showcases a new form of quantum controls in a non-Abelian lattice gauge field.





\begin{acknowledgments}
The authors thank D. Zhai, C. Xiao, H. Zheng for helpful discussions. The work is supported by the National Key R\&D Program of China (No. 2020YFA0309600), National Natural Science Foundation of China (No. 12425406), Research Grant Council of Hong Kong (AoE/P-701/20, HKU SRFS21227S05), and New Cornerstone Science Foundation.
\end{acknowledgments}




\bibliography{Reference}

\end{document}